\documentclass[11pt]{article}
\usepackage{amsmath, amsthm, amsfonts}
\usepackage{amssymb}
\usepackage[margin=1in,footskip=0.25in]{geometry}
\usepackage{mathtools}
\usepackage{makecell}
\usepackage{multicol}
\usepackage{graphicx}
\usepackage{hyperref}
\usepackage{cite}
\theoremstyle{theorem}
\usepackage{lscape}

\newtheoremstyle{defi}
  {10pt}          
  {10pt}  
  {\rm}  
  {\parindent}     
  {\bf}  
  {. }    
  { }    
  {}     
\theoremstyle{defi}

\begin{document}

\date{}

\title{\bf Stability analysis for cosmological models in $f(R)$ gravity using dynamical system analysis}
\author{Parth Shah$^{1}$, Gauranga C. Samanta$^{2}$\\
 $^{1, 2}$Department of Mathematics,
BITS Pilani K K Birla Goa Campus,
Goa-403726, India,\\
parthshah2908@gmail.com \\
gauranga81@gmail.com}

\maketitle

\begin{abstract}
Modified gravity theories have received increased attention lately to understand the late time acceleration of the universe. This viewpoint essentially modifies the geometric components of the universe. Among numerous extension to Einstein's theory of gravity, theories which include higher order curvature invariant, and specifically the class of $f(R)$ theories, have received several acknowledgments. In our current work we try to understand the late time acceleration of the universe by modifying the geometry of the space and using dynamical system analysis. The use of this technique allows to understand the behavior of the universe under several circumstances. Apart from that we study the stability properties of the critical point and acceleration phase of the universe which could then be analyzed with observational data. We consider a particular model  $f(R) = R - \mu R_{c}(R/R_{c})^{p}$ with $ 0 < p < 1, \mu, R_{c} > 0$ for the study. As a first case we consider the matter and radiation component of the universe with an assumption of no interaction between them. Later, as a second case we take matter, radiation and dark energy (cosmological constant) where study on effects of linear, non-linear and no interaction between matter and dark energy is considered and results have been discussed in detail.
\end{abstract}



\section{Introduction}
Late time acceleration predicted by observational data have  opened major challenge in the modern cosmology \cite{sahni,copeland} This causes challenges and questions about limitations of a very successful theory of last century, the general theory of relativity (GR) very prominent. One of the most fruitful approaches so far has been the extended theories of gravity, which have become a standard model in the study of gravitational interaction. These extensions are based on corrections in Einstein's theory. These kind of alternative gravitational theories are an attempt to construct a semi-classical scheme in which GR and most of its successful features can be recovered. This extension essentially consists of adding higher order curvature invariants by modifying the Einstein-Hilbert(EH) action. In the beginning of 1960's there were indications about the merits of this extensions as GR is not renormalizable and thus can not be quantized conventionally. In 1962, Utiyama and De Witt \cite{utiyama} showed that renormalization demands higher order curvature invariants in EH action. These theories created interest among scientific community in higher order theories of gravity, i.e. modifications of EH action to include higher order curvature invariants with respect to Ricci scalar. These corrections to GR were initially considered to be important only at scales close to the Planck scale which is in very early universe and near black hole singularity and indeed there were relevant studies in this attempt like \cite{starobinsky,brandenberger}. However it was not expected that these corrections could give significant effect at low energies i.e. at large scales in the late universe.
Recent evidence from observational physics and cosmology has reveled quite different picture of the universe. The latest CMBR data indicate 4\%, 20 \% and 76\% proportion of ordinary baryonic matter, dark matter and dark energy respectively \cite{astier, eisenstein, riess, spergel}\bibliographystyle{IEEEtran}. The term dark matter refers to an unknown form of matter, which has the clustering properties as ordinary matter but has not yet been discovered experimentally. The term dark energy is an unknown form of energy which is not only discovered experimentally but even does not cluster like ordinary baryonic matter. One could only distinguish them from energy conditions as dark matter satisfy strong energy condition but dark energy does not\cite{wald}. This issue comes with the early time accelerated epoch predicted by inflation which is needed to address the horizon, flatness and monopole problem \cite{kolb, linde, misner, weinberg} as well as to provide a mechanism that generates inhomogeneities which leads to formation of large scale structure \cite{mukhanov}. Apart from that between these two acceleration epoch, there should be a period of decelerated expansion so that radiation and matter dominated eras could take place. Indeed there are observational bounds which suggests that the production of nuclei other that hydrogen takes place in radiation dominated era while matter dominated era is required for the structure formation of the universe \cite{burles, carroll}. These proportions of matter/energy in the universe are surprising and calls for an explanation. The simplest of them all is $\Lambda$CDM supplemented by some scalar field which could lead to inflationary epoch. Besides not explaining the origin of inflation this model has issues like cosmological constant problem, magnitude problem and coincidence problem \cite{carroll1,weinberg1}. These problems make the $\Lambda$CDM model more of an empirical fit to the data without any theoretical motivation. This lead to proposal for other alternatives to dark energy like quintessence \cite{bahcall, caldwell, carroll2, ostriker, peebles, ratra, wang, wetterich} but they do not have strong theoretical motivation behind them. \\
Another way of resolving these issues can be done by arguing that gravity is by far the dominant interacting at cosmological scales and hence, it is the force governing the evolution of the universe. One could attempt to modify the theory of gravity and hence resolve few of the issues of cosmology and astrophysics. It is definitely a different and an acceptable way of approaching for this problem. The precession of Mercury's orbit was attributed initially to some unobserved planet but it took us to the passage from Newtonian gravity to GR and the rest is history.

One of the major problems in theories of gravity is the difficulty in finding out(analytic or numerical) solutions due to highly nonlinear terms in the field equations and hence comparison with observations cannot be carried out easily. So, it is important that other techniques are efficiently used to solve such equations or, at least, to control the overall dynamical behavior. One such method is the Dynamical System Analysis. The approach of this method is to find the
numerical solutions and help in understanding the qualitative behavior of a given physical system \cite{Roy, Odintsov, Odintsov1, Hohmann, Carneiro, Bhatia, Santos, Bamba,  parth}. The most important concept in dynamical system analysis is to find out critical points of a set of first-order ordinary differential equations. The stability conditions are obtained by calculating the Jacobian matrix at critical points and finding their eigenvalues. This is the study of stability properties near a particular critical point. Application of dynamical systems analysis to cosmology has been deeply discussed in these books \cite{Ellis} and \cite{Coley}.    \\

In the present work we try to analyze the stability of the universe which is assumed to have some modification in its geometry part, in particular $f(R)$ gravity. We begin with the viability conditions for this model. This work is overall divided into two major segments. First we consider the mixture of matter and radiation which are assumed to not interact with each other. Later we introduce the dark energy (cosmological constant) along with matter and radiation where the study on effects of linear, non-linear and no interaction has been made. In both the segments, by studying the behaviour, the existence of stability phase and acceleration era has been done with proper calculations and plotting. Section 2 is a brief review and introduction about $f(R)$ theory of gravity, section 3 contains the stability analysis and acceleration phase analysis of the model. Section 4 is about the conclusions and discusses few possible future works.


\section{A Review on $f(R)$ model}
There are numerous ways to modify or deviate from general theory of relativity. Some of the well known alternatives are Scalar Tensor Theory\cite{bergmann, faraoni, nordtvedt, wagoner}, Brans Dicke theory \cite{brans,dicke}, Gauss Bonnet theory \cite{nojiri}, $f(T)$ gravity \cite{bamba1} $f(R,T)$ gravity \cite{harko},  Lovelock gravity \cite{lovelock, padmanabhan} Here we study and give a brief introduction on $f(R)$ gravity which is in detail in \cite{amendola, thomas,salvatore, Nojiri2011, bamba2, motohashi1}. Furthermore, Motohashi et al. \cite{motohashi2} studied a class of viable cosmological models in $f(R)$ gravity and obtained analytic solution for density perturbation in hypergeometric form. Motohashi et al. \cite{motohashi} investigated Phantom boundary crossing and growth index of fluctuations in viable $f(R)$ models, and, based on specific functional form, they calculated numerically the evolution of both homogeneous background and density fluctuations. Subsequently many  authors \cite{Bamba2014, Bamba2015, Nojiri2017, Yousaf2017, Godani, Odintsov2018, Capozziello2018, Odintsovsd} studied cosmological models from various aspects in modified gravity.
This theory comes as a straightforward generalization of the Lagrangian in the Einstein-Hilbert action,
\begin{equation}
S_{EH} = \frac{1}{2 \kappa} \int d^4x \sqrt{-g} R
\end{equation}
where $\kappa = 8\pi G$, $R$ is Ricci Scalar, $g$ is determinant of metric, $g_{\mu\nu}$ = diag$(-1,a^2(t),a^2(t),a^2(t))$ and $a(t)$ is scale factor to become a general function of $R$, i.e.
\begin{equation}{\label{2}}
S = \frac{1}{2 \kappa} \int d^4x \sqrt{-g} f(R)
\end{equation}
$f(R)$ model gives sufficient generality to encapsulate some of the basic characteristics of higher order gravity and yet are rather simple to handle. It is an excellent candidate to be referred as a toy theory i. e. it gives insight of gravity modifications. There are actually two variational principles that can be applied to the action to derive the Einstein's equations. First one being the standard metric variation and other one Palatini variation. This work uses the standard metric variation approach for all discussions.

Beginning with the action in \eqref{2}, by adding the matter term $S_M$, the total action for $f(R)$ takes the form,
\begin{equation}
S = \frac{1}{2 \kappa} \int d^4x \sqrt{-g} f(R) + S_m(g_{\mu \nu}, \psi)
\end{equation}
where $\psi$ denotes the collective matter field.
Variation of this action in standard metric formalism with respect to the metric gives
\begin{equation}{\label{4}}
F(R) R_{\mu \nu} - \frac{1}{2} f(R) g_{\mu \nu} + [ \square g_{\mu \nu} - \nabla_{\mu} \nabla_{\nu}] F(R) = \kappa T_{\mu \nu}
\end{equation}
where, $F(R)$ (also denoted $f_{,R}$) is $\frac{\partial f}{\partial R}$ and as usual,
\begin{equation}{\label{5}}
T_{\mu \nu} = \frac{-2}{\sqrt{-g}}\frac{\delta S_m}{\delta g^{\mu \nu}}
\end{equation}
The trace of equation \eqref{4} is given by
\begin{equation}
3\square F(R) + F(R)R - 2f(R) = \kappa^2 T
\end{equation}

Since 1980s, it was known that the model $f(R) = R + \alpha R^2$ was responsible for inflation in early universe. Inflation ends when quadratic term $R^2$ becomes smaller than linear term $R$. Although this model is not suitable for late time cosmic acceleration. Similarly models like $f(R) = R - \frac{\alpha}{R^n}$ does not satisfy local gravity constrains due to instability. So this makes it very essential to form set of conditions which are viable for $f(R)$ models in metric formalism. These conditions as stated in \cite{amendola1} are:
\begin{itemize}
\item $f_{,R} > 0$ for $R \geq R_{0}$ (where $R_{0}$ is the Ricci scalar at present epoch and is positive). This condition is required to avoid anti-gravity.

\item $f_{,RR} > 0$ for $R \geq R_{0}$. This is required for local gravity tests \cite{dolgov, olmo, faraoni1, navarro} for existence of matter domination era \cite{amendola, amendola2} and for stability of cosmological perturbation \cite{carroll3, song, bean, faulkner}.

\item $f(R) \to R - 2\Lambda$ for $R \gg R_{0}$. This is required for consistency with local gravity tests \cite{amendola3, hu, starobinsky2, appleby, tsujikawa2} and for presence of matter dominated era \cite{amendola}.

\item $0 < \frac{R f_{,RR}}{f_{,R}} < 1$ at $- \frac{R f_{R}}{f} = -2$. This is required for the stability of the late-time de Sitter point \cite{amendola,muller,faraoni2}.
\end{itemize}

\section{Stability Analysis}
In our current work, we do the analysis by considering the model $f(R) = R - \mu R_{c}(R/R_{c})^{p}$ with $ 0 < p < 1, \mu, R_{c} > 0$ \cite{amendola1}. This model satisfies all the local gravity conditions mentioned above and is considered to be a viable model to study the stability analysis of the universe. It is noted here that it satisfies third condition only when value of $p$ is close to 0.
For the flat FLRW space-time the Ricci scalar is given by:
\begin{equation}
R = 6(2H^2 + \dot{H})
\end{equation}
where H is the Hubble parameter. We now construct a model of the universe filled with only matter and radiation and we assume no interaction between them i.e. the usual conservation equations $\dot{\rho_m} + 3H \rho_m =0$ and $\dot{\rho_r} + 4H \rho_r =0$. For this, the explicit form of field equations from equation \eqref{4} are
\begin{equation}\label{8}
\begin{aligned}
3FH^2 = \kappa^2(\rho_m + \rho _r) + \frac{FR - f}{2} - 3H\dot{F} \\
-2F \dot{H} = \kappa^2 (\rho_m + \frac{4}{3} \rho_r) + \ddot{F} - H\dot{F}
\end{aligned}
\end{equation}
Now introducing dimensionless variables,

\begin{equation}
x = - \frac{\dot{F}}{HF}, y = - \frac{f}{6FH^2}, z = \frac{R}{6H^2}, w = \frac{\kappa^2 \rho_r}{3FH^2}
\end{equation}
Without loss of generality, let $\kappa^2 = \frac{8 \pi G}{c^4} = 1$.

Then various density parameters would be,
\begin{equation}
\Omega_r = \frac{\rho_r}{3FH^2} = w, \Omega_m = \frac{\rho_m}{3FH^2} = 1 - x - y - z - w, \Omega_{GC} = x + y + z
\end{equation}

where, $\Omega_{GC}$ represents density parameter due to geometric curvature. From equation \eqref{8}, it is straightforward to derive following set of autonomous differential equations

\begin{align}
x' &= -1 - z - 3y + x^2 - xz + w \\
y' &= \frac{xz}{m} - y(2z - 4 - x) \\
z' &= - \frac{xz}{m}- 2z(z-2) \\
w' &= -2zw + xw
\end{align}

where, prime denotes derivative with respect to $\eta = ln a$ and
\begin{equation}
\begin{aligned}
m \equiv \frac{dlnF}{dlnR} = \frac{R f_{,RR}}{f_{,R}} \\
r \equiv - \frac{dlnf}{dlnR} = - \frac{Rf_{,R}}{f} = \frac{z}{y}
\end{aligned}
\end{equation}
From this, $R$ could be written as a function of $\frac{z}{y}$. We here note that $m$ is a function of $R$, so it follows that $m$ is a function of r, i.e. $m = m(r)$. From the calculations for the model, $f(R) = R - \mu R_{c}(R/R_{c})^{p}$ we deduce
\begin{equation}\label{16}
m = p \left( \frac{r+1}{r} \right) = p \left( \frac{y + z}{z} \right)
\end{equation}

By substituting \eqref{16} in autonomous differential equations, we get
\begin{align}
x' &= -1 - z - 3y + x^2 - xz + w \\
y' &= \frac{xz^2}{p(y+z)} - y(2z - 4 - x) \\
z' &= - \frac{xz^2}{p(y+z)}- 2z(z-2) \\
w' &= -2zw + xw
\end{align}
also,
\begin{equation}
\omega_{eff} = -1 - \frac{2 \dot{H}}{3H^2} = - \frac{1}{3} (2z-1)
\end{equation}

There are 5 real critical points of this system. We will now do the detailed stability and acceleration analysis for all the points.\\

$\mathbf{P_1}: (-4,5,0,0)$. $\Omega_m = 0, \Omega_r = 0, \omega_{eff} = \frac{1}{3}$.
Eigenvalues of this critical point are: -5, -4, 4 and -3. Since one of the eigenvalue is positive, this point can not be stable. Since the value of $\omega_{eff}$ is positive, acceleration for this model is not possible.

$\mathbf{P_2}: (0,-1,2,0)$. $\Omega_m = 0, \Omega_r = 0, \omega_{eff} = -1 $.
Eigenvalues of this critical point are: $-4, -3$, $\frac{-3p - \sqrt{p}\sqrt{-32 + 25p}}{2p} and \frac{-3p + \sqrt{p}\sqrt{-32 + 25p}}{2p}$. We here observe that real part of all the eigenvalues are negative. Hence this point is spiral stable. Apart from that $\omega_{eff}$ is negative, hence this point gives acceleration. This point is completely dominated by geometric curvature as $\Omega_{GC} = 1$. It point can be considered to be regulating the late time expansion of the universe due to finite negative value of $\omega_{eff}$.

$\mathbf{P_3}: \left( \frac{4(p-1)}{p}), -\frac{2(p-1)}{p^2}, \frac{2(p-1)}{p}, \frac{-2+ 8p - 5p^2}{p^2} \right)$. $\Omega_m = 0, \Omega_r = \frac{-2+ 8p - 5p^2}{p^2}, \omega_{eff} = -\frac{1}{3} \left( \frac{3p-4}{p} \right)$.
Eigenvalues of this critical point are: $1, \frac{4 \left(p^3-2 p^2+p\right)}{(p-1) p^2} , \frac{2p + p^3-3 p^2-\sqrt{3} \sqrt{27 p^6-98 p^5+127 p^4-68 p^3+12 p^2}}{2 (p-1) p^2}$,\\ $\frac{2p + p^3-3p^2+\sqrt{3} \sqrt{27 p^6-98 p^5+127 p^4-68 p^3+12 p^2}}{2(p-1)p^2} $. Since one of the eigenvalues is 1, stability will not exist. Also for this model to have acceleration i.e. $\omega < - \frac{1}{3}$, p should be less than 0 or greater than 2 which is not possible.

$\mathbf{P_4}: \left( \frac{3(p-1)}{p}, \frac{3-4p}{2p^2}, \frac{-3 + 4p}{2p},0 \right)$. $\Omega_m = \frac{3-13p-8p^2}{2p^2}, \Omega_r = 0, \omega_{eff} = \frac{1-p}{p}$. Eigenvalues of this critical point are: $\frac{3(p-2p^2+p^3}{(p-1)p^2},-1,-\frac{3}{4p}-p \sqrt{Y}, -\frac{3}{4p}+p \sqrt{Y}$, where $Y = 81-498p+1025p^2-864p^3+256p^4$. When plotted these eigenvalues considering only the real part of third and fourth eigenvalues we note that, this point is spiral stable for the range (0.33,0.71) of $p$. For this model to have acceleration, $p$ should be less than 0 or greater than $\frac{3}{2}$ which is not possible.
\begin{center}
\includegraphics[scale=0.8]{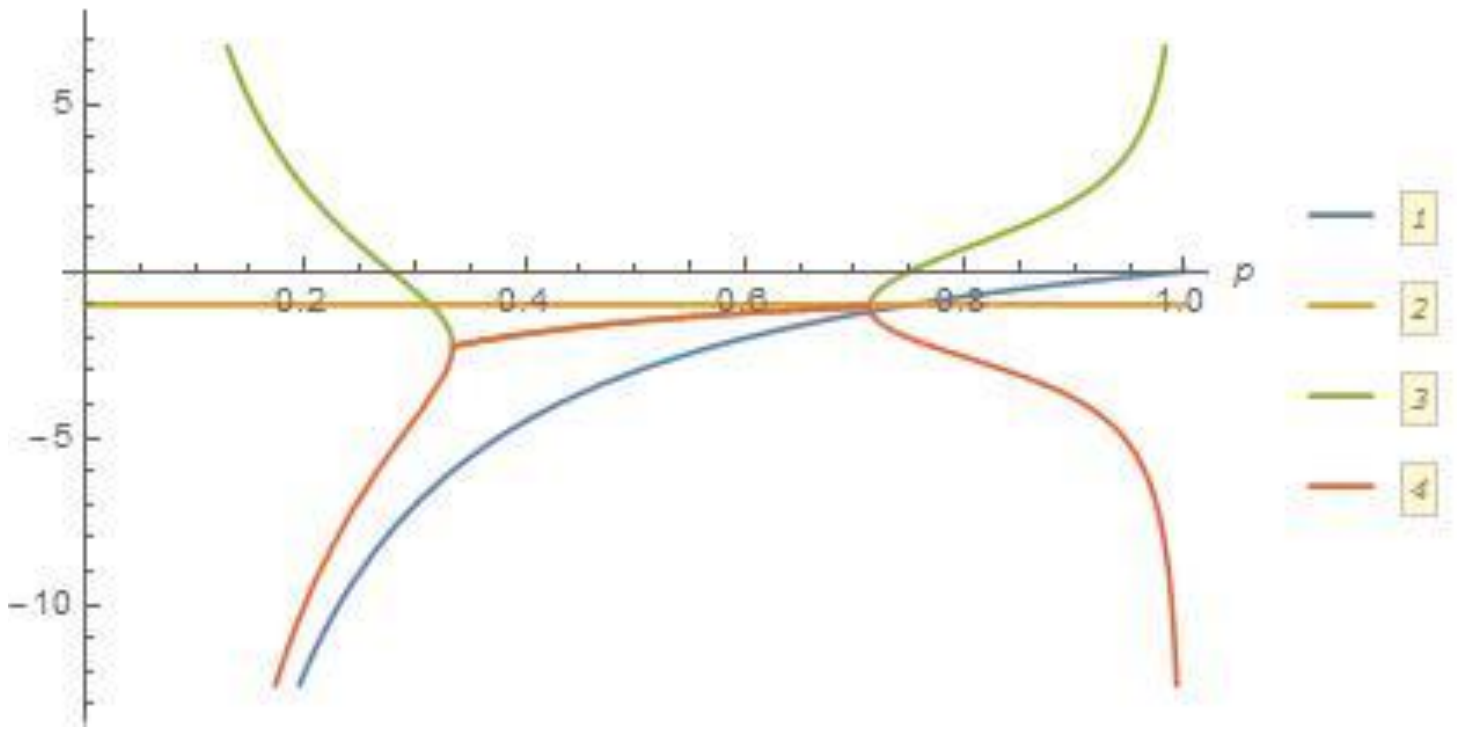}
\end{center}

$\mathbf{P_5}: \left( - \frac{2(p-2)}{2p-1}, \frac{5-4p}{1-3p+2p^2}, \frac{p(4p-5)}{1-3p+2p^2},0 \right)$. $\Omega_m = 0, \Omega_r = 0, \omega_{eff} = -\frac{1}{3} \left[ \frac{6p^2-7p-1}{2p^2-3p+1}\right]$.
Eigenvalues of this critical point are: $- \frac{2(p-2)}{2p-1}, \frac{-8p^2+13p-3}{2p^2-3p+1},\frac{-4p+5}{p-1}, -\frac{2(5p^2-8p+2)}{2p^2-3p+1}$. When these eigenvalues are plotted, we note that for $0 < p < \frac{13 - \sqrt{73}}{16}$ this point is stable. Although, for point to have acceleration, we require $\frac{1}{2} < p < 1$. So this point have stability as well as acceleration but as the range of $p$ is different  we conclude from this point that it represents different eras of the universe. Value of $p$ close to zero where is no acceleration but there is stability could be considered as matter formation as stated in the review. Whereas, value of $p$ close to 1 could be considered as inflation era as $\omega_{eff}$ tends to $-\infty$ and universe is not stable.
\begin{center}
\includegraphics[scale=0.8]{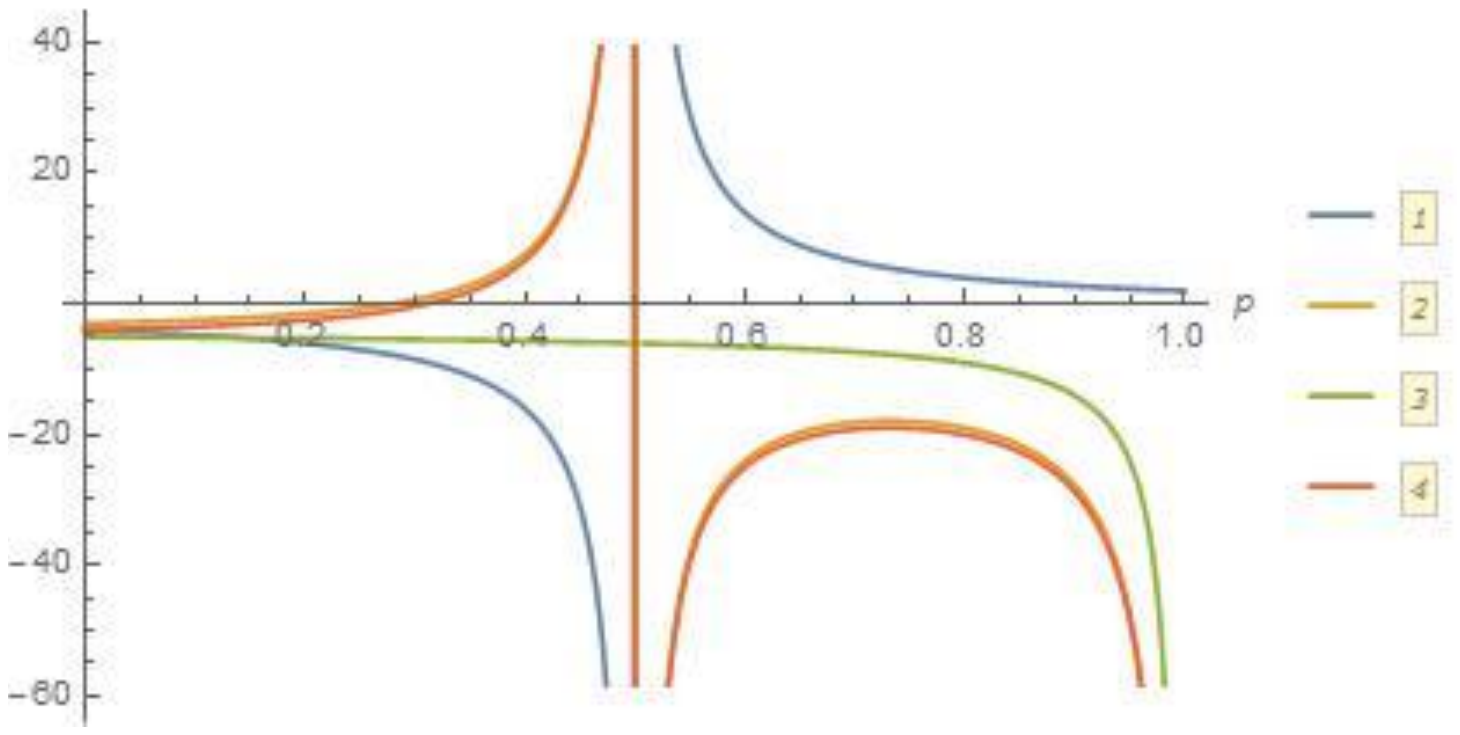}
\end{center}

Now, we include the cosmological constant as a dark energy component in this system. It is further assumed that there is no interaction between any fluid components.
Then the field equations then would be:
\begin{equation}\label{22}
\begin{aligned}
3FH^2 = \kappa^2(\rho_m + \rho _r + \rho_{\Lambda}) + \frac{FR - f}{2} - 3H\dot{F} \\
-2F \dot{H} = \kappa^2 (\rho_m + \frac{4}{3} \rho_r) + \ddot{F} - H\dot{F}
\end{aligned}
\end{equation}

This needs an introduction one extra dimensionless variable, which then gives,
\begin{equation}
x = - \frac{\dot{H}}{HF}, y = - \frac{f}{6FH^2}, z = \frac{R}{6H^2}, w = \frac{\rho_r}{3FH^2}, s = \frac{\rho_m}{3FH^2}
\end{equation}

Then density parameter for matter and cosmological constant would be,
\begin{equation}
\Omega_m = \frac{\rho_m}{3FH^2} = s, \Omega_{\Lambda} = 1- (x + y + z + w + s)
\end{equation}

From equation \eqref{22}, it is straightforward to derive following set of autonomous differential equations and substituting $m$ from \eqref{16}

\begin{align}
x' &= -4 + 5x + 2z + 4w + 3s - xz + x^2 \\
y' &= \frac{xz^2}{p(y+z)} - y(2z - 4 - x) \\
z' &= - \frac{xz^2}{p(y+z)}- 2z(z-2) \\
w' &= -2zw + xw \\
s' &= -2zs + sx + s
\end{align}

There are 3 real critical points of this system. Their detailed stability and acceleration analysis would be done. The eigenvalues of each of these points are of very high order and analysis is shown by plots. \\

$\mathbf{P_6}: \left( \frac{4(p-1)}{p},-\frac{2(p-1)}{p^2}, \frac{2(p-1)}{p},\frac{-2+10p-7p^2}{p^2},0 \right)$. $\Omega_m = 0, \Omega_r = \frac{-2+10p-7p^2}{p^2}, \Omega_{\Lambda}= 2 - \frac{2}{p}, \omega_{eff} = -\frac{1}{3} \left( \frac{3p-4}{p}\right)$. The acceleration for this model happens only when $p < 0$ or $p > 2$, which is not possible as per our definition. Also value of one of the eigenvalues is 1 and this point is not stable. The plot for real part of the eigenvalues looks like
\begin{center}
\includegraphics[scale=0.8]{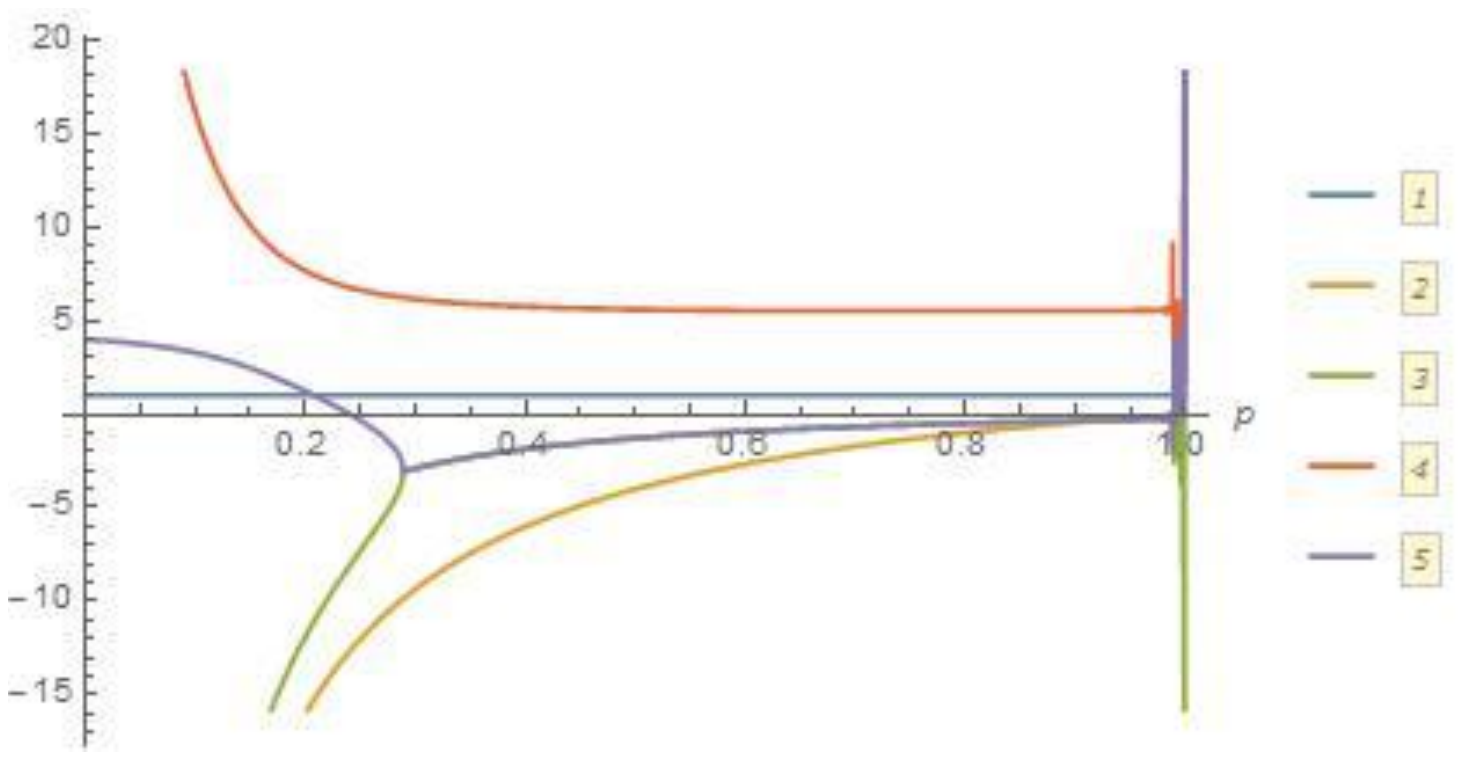}
\end{center}

$\mathbf{P_7}: \left( \frac{2(p-1)}{p},-\frac{3-4p}{2p^2}, \frac{-3+4p}{2p},0,\frac{-3+17p-12p^2}{2p^2} \right)$. $\Omega_m = \frac{-3+17p-12p^2}{2p^2}, \Omega_r = 0, \Omega_{\Lambda}= 2 - \frac{2}{p}, \omega_{eff} = -\frac{1}{3} \left( \frac{3p-4}{p}\right)$. The acceleration for this model happens only when $p < 0$ or $p > \frac{3}{2}$, which is not possible as per our definition. Plot of the eigenvalues makes it trivial that stability is not possible in this case.
\begin{center}
\includegraphics[scale=0.8]{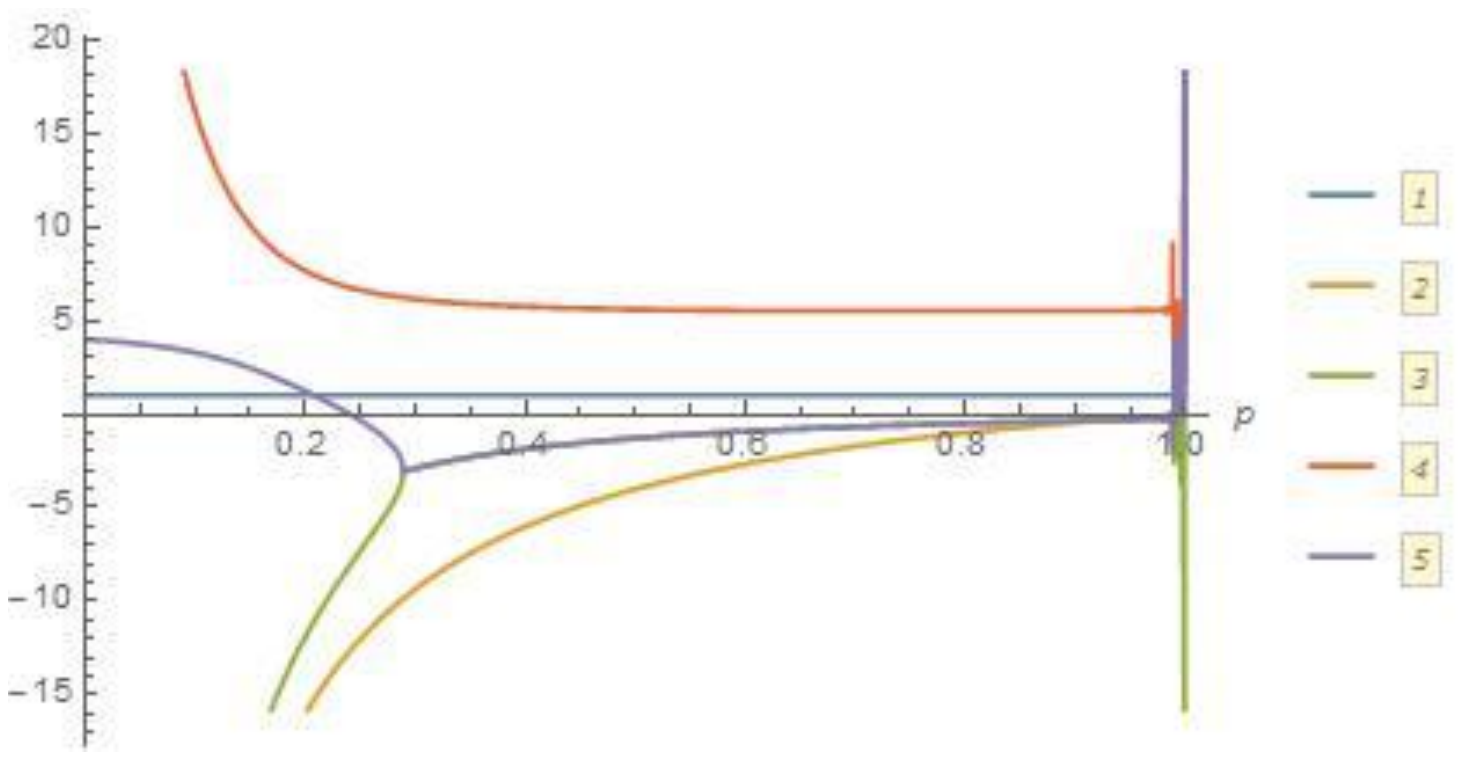}
\end{center}

$\mathbf{P_8}: \left( \frac{-2(3p-4)}{2p-1},\frac{2+3p-4p^2}{p(1-3p+2p^2)}, \frac{-2-3p+4p^2}{1-3p+2p^2},0,0 \right)$. $\Omega_m = 0, \Omega_r = 0, \Omega_{\Lambda}= 2 - \frac{2}{p}, \omega_{eff} = -\frac{1}{3} \left( \frac{6p^2-3p-5}{2p^2-3p+1}\right)$. The acceleration for this model happens only when $\frac{1}{2} < p < 1$.  Plot of the eigenvalues makes it trivial that stability is not possible in this case. So here acceleration is possible but it is not stable.
\begin{center}
\includegraphics[scale=0.8]{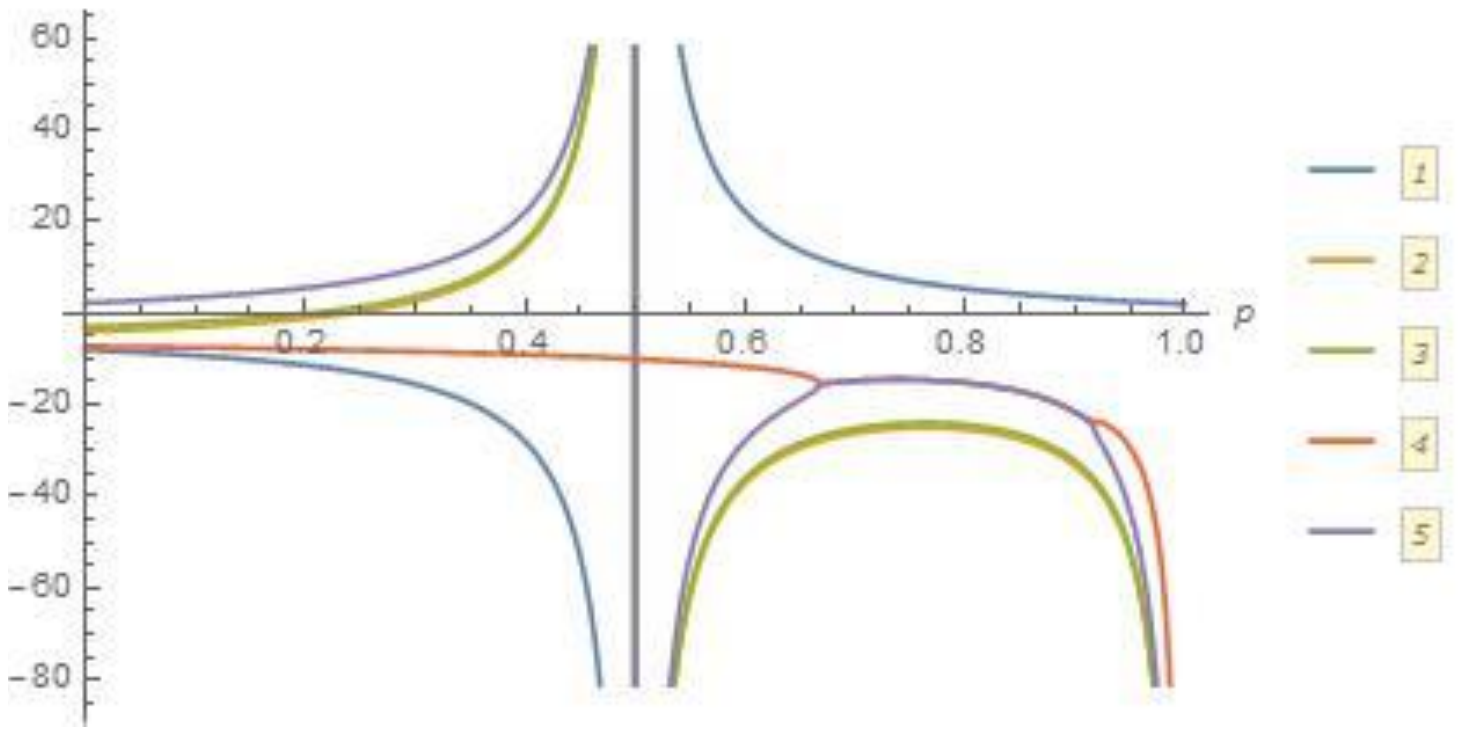}
\end{center}

Now, we investigate the impacts of interaction between matter and dark energy. In current work, both linear as well as non-linear interactions have been considered. For interaction, the dimensionless variables have not been tickled. Taking interaction Q between matter and dark energy, the continuity equations read,
\begin{equation}
\dot{\rho_r} + 4H \rho_r =0, \dot{\rho_m} + 3H \rho_m = Q \text{ and } \dot{\rho_{\Lambda}} = -Q
\end{equation}

Case I: Linear Interaction $(Q = H\rho_{tot})$ \cite{golchin}
The set of autonomous differential equation in the case of linear interaction stated above would be:

\begin{align}
x' &= -4 + 5x + 2z + 4w + 3s - xz + x^2 \\
y' &= \frac{xz^2}{p(y+z)} - y(2z - 4 - x) \\
z' &= - \frac{xz^2}{p(y+z)}- 2z(z-2) \\
w' &= -2zw + xw \\
s' &= 1-x-y-z+s-2zs+sx
\end{align}

There are 3 real critical points of this system. The detailed stability and acceleration analysis would be done. The eigenvalues of each of these points are of very high order and analysis is shown by plots. \\

$\mathbf{P_9}: \left( -4,-3,0,0,\frac{8}{3} \right)$. $\Omega_m = \frac{8}{3}, \Omega_r = 0, \Omega_{\Lambda}= \frac{16}{3}, \omega_{eff} = \frac{1}{3}$. The acceleration for this point is not possible. The Eigenvalues of this point are $\left(-2-\frac{\left(1-i \sqrt{3}\right) \sqrt[3]{\frac{1}{2} \left(9+i \sqrt{6063}\right)}}{2\ 3^{2/3}}-\frac{4 \left(1+i
   \sqrt{3}\right)}{\sqrt[3]{\frac{3}{2} \left(9+i \sqrt{6063}\right)}} \right),-4,4$, $\left(-2-\frac{\left(1+i \sqrt{3}\right) \sqrt[3]{\frac{1}{2} \left(9+i
   \sqrt{6063}\right)}}{2\ 3^{2/3}}-\frac{4 \left(1-i \sqrt{3}\right)}{\sqrt[3]{\frac{3}{2} \left(9+i \sqrt{6063}\right)}} \right), \left( -2+\frac{\sqrt[3]{\frac{1}{2}
   \left(9+i \sqrt{6063}\right)}}{3^{2/3}}+\frac{8}{\sqrt[3]{\frac{3}{2} \left(9+i \sqrt{6063}\right)}} \right)$. Clearly this point is not stable as one eigenvalue is 4.

$\mathbf{P_{10}}: \left(0,-1,2,0,0 \right)$. $\Omega_m = 0, \Omega_r = 0, \Omega_{\Lambda}= 0, \omega_{eff} = - 1$. This point does provide acceleration phase. In the plot, real part of eigenvalues till $p=10$ is shown due to overlap in the earlier value. It is evident that this point is spiral stable as all eigenvalues have negative real part in the region $0 < p < 1$.
\begin{center}
\includegraphics[scale=0.8]{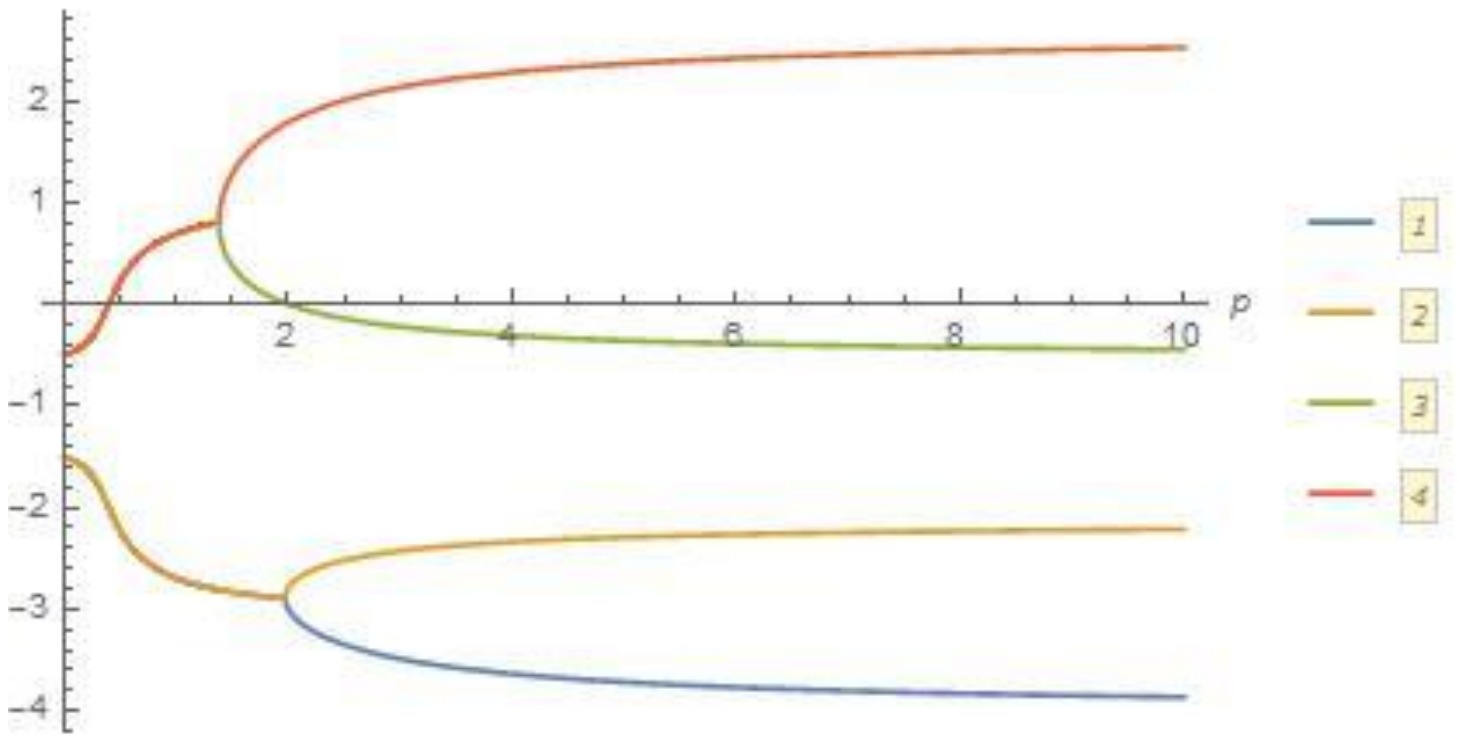}
\end{center}

$\mathbf{P_{11}}: \left( \frac{4(p-1)}{p}, - \frac{2(p-1)}{p^2}, \frac{2(p-1)}{p}, \frac{-14+64p-43p^2}{4p^2},\frac{2-8p+5p^2}{p^2} \right)$. $\Omega_m = \frac{2-8p+5p^2}{p^2}, \Omega_r = \frac{-14+64p-43p^2}{4p^2}, \Omega_{\Lambda}= \frac{7}{2 p^2}-\frac{4}{p}+\frac{3}{4}, \omega_{eff} = - \frac{1}{3} \left(\frac{3p-2}{p} \right)$. This point provide acceleration for $p < 0$ or $p > 1$, which is not possible as per the definition. This point is also unstable from the plot of real part of eigenvalues.
\begin{center}
\includegraphics[scale=0.8]{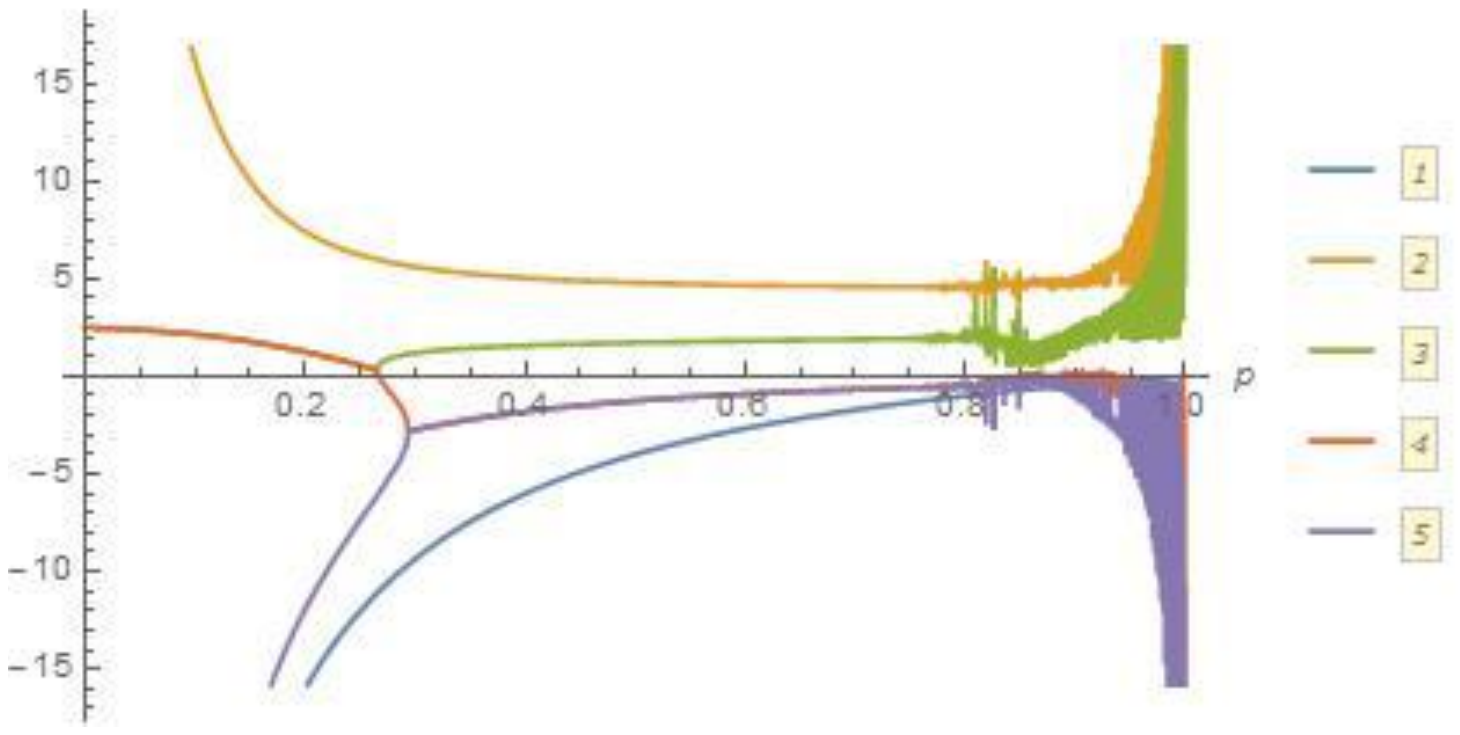}
\end{center}

Case II: Non-Linear Interaction $(Q = H\frac{\rho_{\Lambda}\rho_{m}}{\rho_{tot}})$ \cite{golchin} The set of autonomous differential equations for the non-linear interaction would be:
\begin{align}
x' &= -4 + 5x + 2z + 4w + 3s - xz + x^2 \\
y' &= \frac{xz^2}{p(y+z)} - y(2z - 4 - x) \\
z' &= - \frac{xz^2}{p(y+z)}- 2z(z-2) \\
w' &= -2zw + xw \\
s' &= \frac{s(1-y-3z-w-x^2+2z^2+xz+2yz-xy-sx-sy-sz)}{1-x-y-z}
\end{align}

There are 4 real critical points of this system. The detailed stability and acceleration analysis would be done. Again like the previous case, the eigenvalues of each of these points are of very high order and analysis is shown by plots. \\

$\mathbf{P_{12}}:(-4, 13, 0, 0, \frac{8}{3})$. $\Omega_m = \frac{8}{3}, \Omega_r = 0, \Omega_{\Lambda}= - \frac{32}{3}, \omega_{eff} = \frac{1}{3}$. The acceleration for this point is not possible. The Eigenvalues of this point are $\left(-\frac{25 \left(1+i \sqrt{3}\right)}{3 \sqrt[3]{\frac{1}{2} \left(-117+i \sqrt{486311}\right)}}-\frac{1}{6} \left(1 \pm i \sqrt{3}\right)
   \sqrt[3]{\frac{1}{2} \left(-117+i \sqrt{486311}\right)} \right)$, $-4,4$, $ \left( \frac{1}{3} \left(\frac{50}{\sqrt[3]{\frac{1}{2} \left(-117+i
   \sqrt{486311}\right)}}+\sqrt[3]{\frac{1}{2} \left(-117+i \sqrt{486311}\right)}\right) \right)$. Since one eigenvalue is positive, stability is not possible. \\

$\mathbf{P_{13}}: \left( \frac{4(p-1)}{p},-\frac{2(p-1)}{p^2}, \frac{2(p-1)}{p},\frac{-2+10p-7p^2}{p^2},0 \right)$. $\Omega_m = 0, \Omega_r = \frac{-2+10p-7p^2}{p^2}, \Omega_{\Lambda}= 2 - \frac{2}{p}, \omega_{eff} = -\frac{1}{3} \left( \frac{3p-4}{p}\right)$. The acceleration for this model happens only when $p < 0$ or $p > 2$, which is not possible as per our definition. Also value of one of the eigenvalues is 1 and this point is not stable. The plot for real part of the eigenvalues looks like
\begin{center}
\includegraphics[scale=0.8]{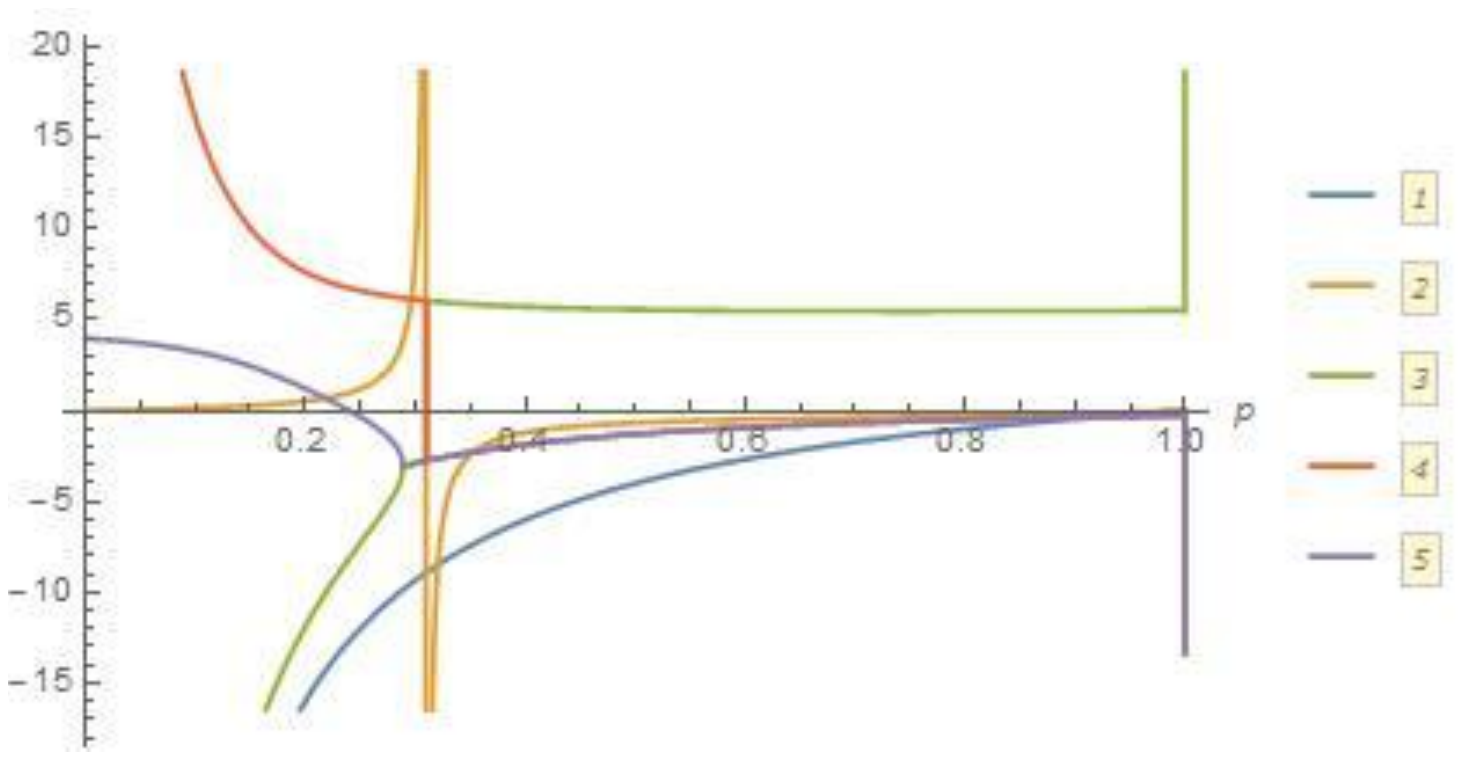}
\end{center}

$\mathbf{P_{14}}: \left( \frac{4 (p-1)}{p},\frac{2 (p-1)}{p^2},\frac{2 (p-1)}{p},\frac{-153 p^4+440 p^3-418 p^2+144 p-16}{p^2 \left(21 p^2-32 p+8\right)},\frac{8\left(p^2-p\right)}{21 p^2-32 p+8}\right)$.  $\Omega_m = \frac{8\left(p^2-p\right)}{21 p^2-32 p+8}, \Omega_r = \frac{-153 p^4+440 p^3-418 p^2+144 p-16}{p^2 \left(21 p^2-32 p+8\right)}, \Omega_{\Lambda}= \frac{4 \left(10 p^4-47 p^3+73 p^2-44 p+8\right)}{p^2 \left(21 p^2-32 p+8\right)}, \omega_{eff} = -\frac{1}{3} \left( \frac{3p-4}{p}\right)$. So acceleration occurs only when $p<0$ or $p>2$, which is not possible. Plot of eigenvalues indicate that atleast one eigenvalue is negative for all values in the region $0<p<1$. So stability is not possible.
\begin{center}
\includegraphics[scale=0.8]{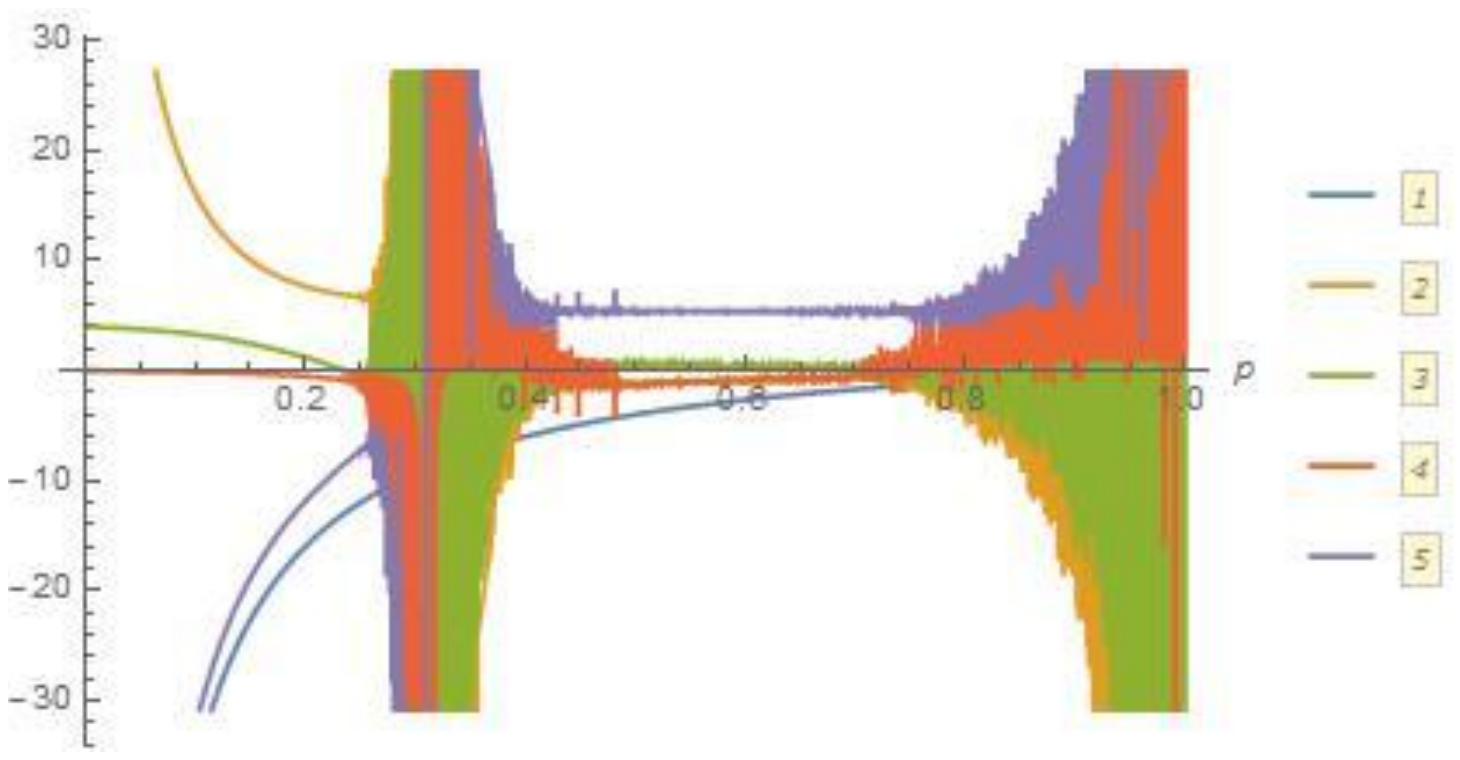}
\end{center}

$\mathbf{P_{15}}: \left( \frac{-2(3p-4)}{2p-1},\frac{2+3p-4p^2}{p(1-3p+2p^2)}, \frac{-2-3p+4p^2}{1-3p+2p^2},0,0 \right)$. $\Omega_m = 0, \Omega_r = 0, \Omega_{\Lambda}= 2 - \frac{2}{p}, \omega_{eff} = -\frac{1}{3} \left( \frac{6p^2-3p-5}{2p^2-3p+1}\right)$. The acceleration for this model happens only when $\frac{1}{2} < p < 1$.  Plot of the eigenvalues makes it trivial that stability is not possible in this case. So here acceleration is possible but it is not stable.
\begin{center}
\includegraphics[scale=0.8]{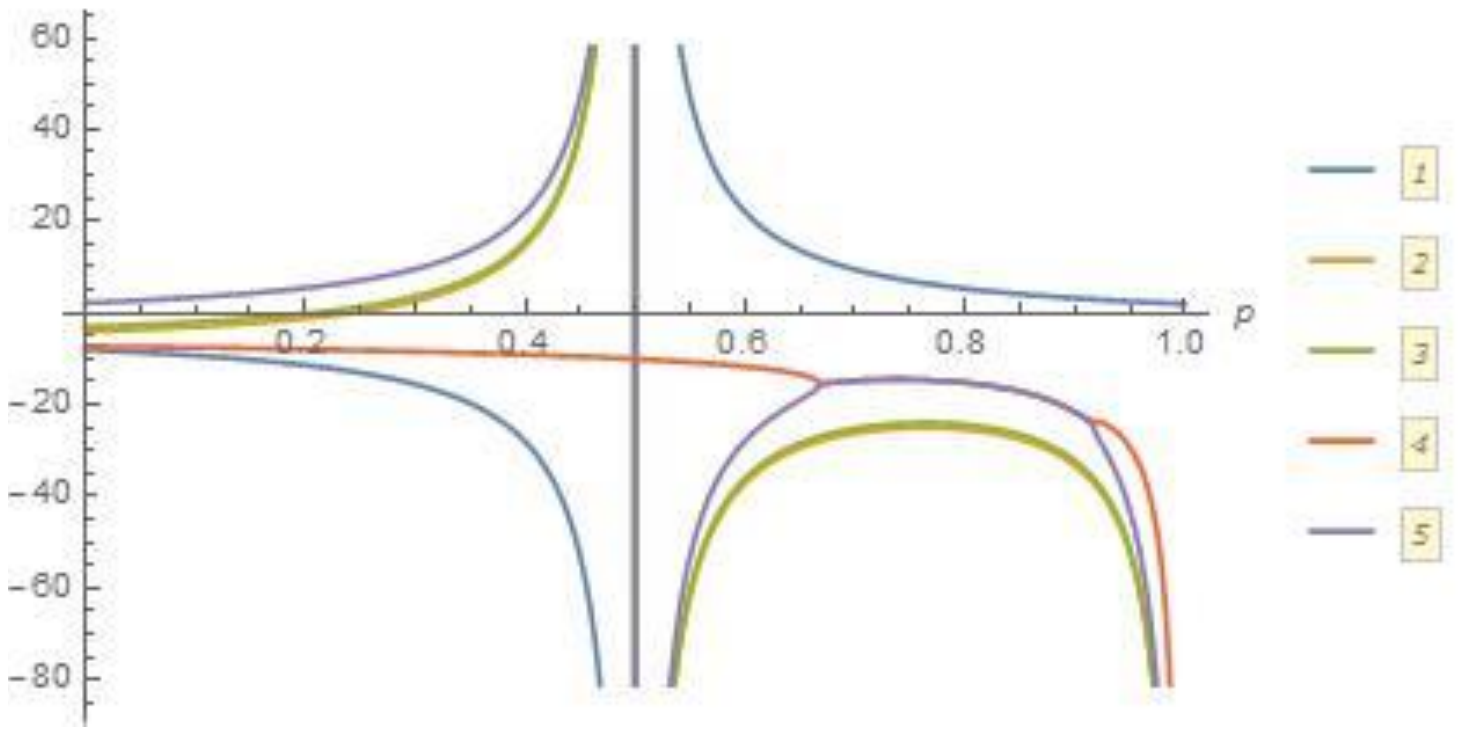}
\end{center}

\section{Conclusion and Further work}
In this work, an extensive literature survey on modified theories of gravity has been studied. The issues about Einstein's general theory of relativity have been looked and discussed briefly. This leads into consideration of modified theory of gravity and this work extensively uses $f(R)$ model, which is essentially the modification of the geometry part of the universe. An overall discussion about the $f(R)$ gravity in metric is beginning from its action principle to field equations in standard metric formalism. After that a brief discussion on certain viability conditions regarding the type of model to be chosen is done. This specific choice of function $f(R)$ lead to fully realistic model which could be compared in detail with cosmological observations. \\
The analysis part of this work is fully devoted to one of the particular function $f(R) = R - \mu R_{c}(R/R_{c})^{p}$ with $ 0 < p < 1, \mu, R_{c} > 0$.       The method of dynamical system analysis is being used as we introduce dimensionless variables from the corresponding field equations. This lead to formation of system of autonomous differential equations corresponding to the function $f(R)$. We begin with the universe filled only with matter and radiation which do not interact with each other. All of the real critical points of the system are evaluated.  Moreover calculation of $\omega_{eff}$ which gives the acceleration phase and signature of the eigenvalues of the Jacobian matrix of the corresponding eigenvalues which gives the stability analysis are also done. This entire problem is in 4 dimension. The brief analysis of the result is as follows:
\begin{center}
\begin{tabular}{ |c|c|c| }
 \hline \textbf{Point} & \textbf{Stability} & \textbf{Acceleration}\\
 \hline
 $P_1$ & Not Stable & Never \\
 \hline
 $P_2$ &  Spiral Stable & Always \\
 \hline
 $P_3$ &  Not stable & Never \\
 \hline
 $P_4$ & Spiral stable for $p \in (0.33,0.71)$ & Never \\
 \hline
 $P_5$ & Stable of $0 < p < \frac{13 - \sqrt{73}}{16}$ & $\frac{1}{2} < p <
1$ \\
  \hline
\end{tabular}
\end{center}

We further take this problem to fifth dimension by introducing dark energy (cosmological constant) along with matter and radiation. Similar calculation about evaluating acceleration phase with the help of $\omega_{eff}$ and stability analysis with the help of signature of eigenvalues of Jacobian matrix of corresponding eigenvalues is done. We assume that there is no interaction between these cosmological fluids. Brief summary of result is as follows.
\begin{center}
\begin{tabular}{ |c|c|c| }
 \hline \textbf{Point} & \textbf{Stability} & \textbf{Acceleration}\\
 \hline
 $P_6$ & Not Stable & Never \\
 \hline
 $P_7$ &  Not Stable & Never \\
 \hline
 $P_8$ &  Not stable & $0< p < \frac{1}{2}$ \\
 \hline
\end{tabular}
\end{center}

Now, we take two subcases in the mixture of matter, radiation and dark energy (cosmological constant). One is with the linear interaction ($Q = H\rho_{tot}$) and other is the non-linear ($Q=H \frac{\rho_{\Lambda} \rho_{m}}{\rho_{tot}}$) interaction between matter and dark energy. Similar analysis for both the kinds of interaction have been done and summary of results is as follows: \\

\textbf{For linear interaction:}
\begin{center}
\begin{tabular}{ |c|c|c| }
 \hline \textbf{Point} & \textbf{Stability} & \textbf{Acceleration}\\
 \hline
 $P_9$ & Not Stable & Never \\
 \hline
 $P_{10}$ &  Stable Spiral & Always \\
 \hline
 $P_{11}$ &  Not Stable & Never \\
 \hline
\end{tabular}
\end{center}

\textbf{For non-linear interaction:}
\begin{center}
\begin{tabular}{ |c|c|c| }
 \hline \textbf{Point} & \textbf{Stability} & \textbf{Acceleration}\\
 \hline
 $P_{12}$ & Not Stable & Never \\
 \hline
 $P_{13}$ &  Not Stable & Never \\
 \hline
 $P_{14}$ &  Not Stable & Never \\
 \hline
 $P_{15}$ &  Not Stable & $0 < p < \frac{1}{2}$ \\
 \hline
\end{tabular}
\end{center}

Here all the, $P_1, P_2,$ etc are the points as stated in the calculation part. As eigenvalues in most of cases are complicated, plots of behaviour of eigenvalue against $p$ in the range $0 < p < 1$ is made. This gives the clear idea of the stability analysis. From all the above calculations and discussions, we could conclude that in the case of modified gravity (here $f(R)$) there is no effect of the use of cosmological constant component. Stability and acceleration phase are achieved without adding this 'unobserved' quantity. Point $P_{10}$ which is under linear interaction is stable spiral and also shows acceleration, but if we observe it closely the value of $\Omega_{\Lambda}$ of that point is 0. This show that it does not contain any dark energy component (cosmological constant). This work is motivated to theoretically prove accelerated expansion of the universe along with stability, matter formation era and acceleration phase. Point $P_{5}$ shows the deceleration phase for small values of $p$, which as mentioned in the literature is essential of presence of matter formation epoch. It signifies the beginning of that era. This same point gives the inflationary acceleration at values of $p$ close to 1 as it gives a very large negative value of $\omega_{eff}$. Point $P_{2}$ gives us the late time acceleration as it has stability and $\omega_{eff}$ is negative and finite. Similarly point $P_4$ signifies the beginning of radiation domination epoch. \\
This work could be extended by considering different $f(R)$ model which satisfies all the viability conditions. Apart from this some different category of modification like $f(R,T)$ or Gauss Bonnet theory could be considered and the use of dynamical system analysis could be applied further and stability analysis for such cases could be carried out.

\section{Acknowledgement}
The authors are very much thankful to the reviewers for their constructive comment to improve the quality of work.  The author Parth Shah is extremely thankful to Department of Science and Technology (DST), Govt. of India, for providing INSPIRE Fellowship (Ref. No. IF160358) for carrying out his research work and Department of Mathematics, BITS Pilani Goa campus for providing other research facilities.

\end{document}